\documentstyle[prl,aps,epsfig]{revtex}
\input{psfig}
\begin{document}
\title{Molecular Dynamics Studies on HIV-1 Protease: Drug Resistance and 
Folding Pathways.}
\author{Fabio Cecconi$^1$, Cristian Micheletti$^1$, Paolo
Carloni$^{1,2}$ and Amos Maritan$^1$} 

\address{(1) International
School for Advanced Studies (SISSA/ISAS), Via Beirut 2-4, I-34014
Trieste,\\ INFM and ``The Abdus Salam'' International Centre for
Theoretical Physics.}  \address{(2) International Centre for Genetic
Engineering and Biotechnology (ICGEB) - AREA Science Park, Padriciano
99, I-34012 Trieste} \date{\today} \maketitle

\tighten
\begin{abstract}
Drug resistance to HIV-1 Protease involves accumulation of multiple
mutations in the protein. Here we investigate the role of these
mutations by using molecular dynamics simulations which exploit the
influence of the native-state topology in the folding process. Our
calculations show that sites contributing to phenotypic resistance of
FDA-approved drugs are among the most sensitive positions for the
stability of partially folded states and should play a relevant role
in the folding process. Furthermore, associations between amino acid
sites mutating under drug treatment are shown to be statistically
correlated.  The striking correlation between clinical data and our
calculations suggest a novel approach to the design of drugs tailored
to bind regions crucial not only for protein function but also for
folding.
\end{abstract}

\vskip 2.0cm

\section{Introduction}

The human immunodeficiency virus encodes a protease (HIV-1 PR)
cleaving the gag and the gag-pol viral polyproteins into enzymes and
structural proteins \cite{gulnik}. The discovery that inhibition of
this protease (a homodimer of 198 amino acids) causes the formation of
non-infectious virus particle has prompted an enormous effort to
design efficient inhibitors against AIDS attack \cite{hiv1}.
Currently, five antiviral agents are approved by FDA: Saquinavir
(SQV), Ritonavir (RTN), Indinavir (IND), Nelfinavir (NLF), Amprenavir
(APR) and several others are under clinical trials
\cite{gulnik,hiv1,apr,BIOCH88}. Therapeutic benefit is unfortunately
short-lived as the virus strains - evolving under drugs' selective
pressure - encode HIV-1 PR multiple mutants with low drugs affinity
\cite{gulnik,BIOCH88,condra,condra1,boucher,boucher1}: mutants
resistant to protease inhibitors can emerge in vivo already after less
than one year \cite{condra}. The problem of drug-resistance persists
also when a combination of PR and reverse transcriptase (RT) inhibitor
are used.

The occurrence of mutations withstanding antiviral drugs is not a mere
consequence of drug action, rather it results from viral
replication itself \cite{condra}. Indeed, mutations are found rather
irrespective of drug structural diversity involving virtually every
protein domain: this is the case for HIV-1 PR mutants which are
resistant against FDA approved drugs \cite{BIOCH88}. Among such
mutations, a few involve the active site (residue 25 - aspartic acid
\cite{condra1}) while the others belong to protein regions away from
it.

By means of molecular dynamics (MD) simulations within the framework
introduced in Ref.~\cite{Micheletti99b} (see next section), we show
here that the positions where these mutations occur play a key role in
the proximity of temperatures where the specific heat peaks occur.
Moreover, we shall argue that residues involved in a frequently
observed covariant mutation \cite{condra}, are statistically
correlated. Indeed, while the study of the native state structure can
lead to a rational design of drugs binding the active site (or
otherwise disrupt the biological function of the agent by acting on
its native structure) the analysis of the folding pathways can provide
fundamental information \cite{Fersht}. In particular, it can reveal,
as in the case of HIV-1 PR, the presence of kinetic bottlenecks
associated to severe entropy reduction that inhibits the progress
toward the native state. These bottlenecks represent the most delicate
part of the folding process. They are followed by the sudden formation
of specific native-like protein sub-regions and afterwards the folding
process proceeds rapidly until another bottleneck is 
reached. The identification of sites involved in the bottlenecks and
their correlation with the active site is crucial pharmaceutically
because they are the ideal targets of effective drugs. From this point
of view, due to the large amount of data available on drug resistance,
HIV-1 PR is an excellent candidate to validate our automatic strategy
to identify key folding sites.  In the following, we shall present
evidence showing that the crucial sites can be identified
with good statistical confidence. The framework introduced here is
general and, applied to other viral proteins, ought to be useful for
suggesting which sites should be preferentially targeted by effective
drugs.

\section{Theory}

The strategy adopted here to identify the crucial sites for the folding and
assembling of HIV-1 PR is based on a recent theoretical framework
\cite{Micheletti99b,M00} that allows to capture the main features of
the folding process by a simplified description of both the protein
structure and the folding dynamics.
At the basis of the method is the observation that the topology of the
native state plays a crucial role in steering the folding process
\cite{Micheletti99b,Finkel,eaton,baker,dobson,serrano,Clem}. This
statement is supported by an increasing amount of experimental
evidence. Perhaps, the most notable examples are: (a) the close
similarity of the transition-state conformations of proteins having
structurally-related native states (despite the very poor sequence
similarity) \cite{sh3A,sh3B} and (b) the strong influence that certain
simple topological properties, such as the contact order, have on
protein folding rates \cite{Plaxco}. Such observations, and others
summarised in the recent review of D.~Baker \cite{bakernature},
~complement the findings of Anfinsen, who established that a protein's
amino acid sequence uniquely encodes its native state
\cite{Anfinsen73}.  Indeed, since the topology of the native state
influences the folding process, the amino acid sequence must also
encode its possible folding pathways.

We focus our attention on the topological
rate-limiting steps along the pathways from unfolded states to the
native one. Such bottleneck stages, are usually found in
correspondence of non-local amino acid interactions that require the
overcoming of a large entropy barrier (due to the flexibility of the
peptide chain intervening between them); the formation of such crucial
contacts acts as a nucleus for the establishment of further native
interactions and leads to a rapid progress along the folding reaction
coordinate until another barrier is met.

It is striking that the sites involved in the topological bottlenecks
are those where the largest changes in the folding kinetics are
observed in site-directed mutagenesis experiments \cite{Fersht}, as
first established for CI2 and Barnase \cite{Micheletti99b}. This shows
that nature has carefully optimised the protein sequence so to exploit
the conformational entropy reduction accompanying the folding process
\cite{funnel} through the careful choice of the amino acids forming
the crucial contacts.

With the purpose of identifying the key sites we investigate the
topological obstacles encountered during the formation of the native
HIV-1 PR structure. Such sites are, intuitively, the ideal candidate
targets of effective drugs, as they take part to the most delicate
steps of the folding process. This fact was first recognised by
Anfinsen in connection with the staphylococcal nuclease
\cite{Anfinsen73}. The most effective strategy to prevent the protein
formation is acting on residues involved in the key contacts and
undermining the formation/overcoming of bottleneck stages. One of the
distinctive features of the HIV virus is the extremely high rate of
mutations. The capability of encoding several mutants provides a
possibility for HIV-1 PR to elude the disruptive action of the drug by
intervening on the key sites. This seems an unavoidable countermeasure
since the viable mutants (i.e. those with native-like enzymatic
activity) retain the original native structure and hence, arguably,
encounter the same bottlenecks as the wild-type. Within this
framework, the lapse of time during which the drug therapy is
temporarily effective, corresponds to the time taken by the virus to
encode, through random mutations, a mutant form of HIV-1 PR where the
crucial sites have been fine-tuned to overcome not only the kinetic
bottlenecks (as for the wild type) but also the additional drug
attack. The key sites identified through the method explained in the
next sections, have been compared with the known key mutating
positions of HIV-1 PR, finding a highly significant correlation
between the two of them. In addition, previously unexplained
co-variant mutations seen in HIV-1 PR are explained as arising due to
the correlation between distinct topological bottlenecks.

\section{Methods}
	
The model that we adopted encompasses an energy-scoring function of
the Go-type \cite{Go}.  This is one of the simplest energy functionals
and provides a natural topological bias to the native state  
by rewarding the formation of native pairwise interactions. In the
version used here, which is a generalization of
ref. \cite{Micheletti99b} apt for molecular dynamics studies, the
cooperativity of the folding process is enhanced by the introduction
of repulsive non-native interactions. In our Hamiltonian, each pair of
non-consecutive amino acids interacts with the following strength:

\begin{equation}
5 V_0 \varepsilon_{ij}^N \left[ 
\left( \frac{r^N_{ij}}{r_{ij}} \right)^{12} - \frac{6}{5} 
\left( \frac{r^N_{ij}}{r_{ij}} \right)^{10} \right] + 
V_1 (1-\varepsilon^N_{ij}) \left(\frac{r_0}{r_{ij}}\right)^{12},
\label{eq:inter}
\end{equation} 

\noindent where $r_0 = 6.8$\AA, $r^N_{i,j}$ denotes the distance of
$C_\alpha$ atoms of amino acids $i$ and $j$ in the native structure
and $\varepsilon^N_{ij}$ is the native contact map, whose entries are
1 (0) if $i$ and $j$ are (not) in contact in the native conformation
(i.e. below or above 6.5 \AA). $V_0$ and $V_1$ are constants
controlling the strength of interactions ($V_0 = 20$, $V_1=V_0/400$ in
our simulations). In addition, the peptide bond between two
consecutive amino acids, $i$, $i+1$ at distance $r_{i,i+1}$ is
described by the unharmonic potential:
\begin{equation}
\frac{a}{2} (r_{i,i+1} - r_d)^2 + \frac{b}{4} (r_{i,i+1} - r_d)^4
\end{equation}
with parameters $a = V_0$, $b = 10 V_0$, and 
$r_d =3.8$ \AA is the rest distance between consecutive $C_{\alpha}$ atoms.  

It is important to notice that in Eq.~(\ref{eq:inter}) the formation
of any native contact is rewarded in the same way, since $V_0$ does
not depend on $i$ and $j$.  This choice is done deliberately, so that
the only information entering Eq.~(\ref{eq:inter}) is the native
contact map and not the types of interacting amino acids (i.e. no
sequence information).  While by construction, the minimum of the
energy scoring function is achieved in correspondence of the native
state, there is no {\em a priori} guarantee that the folding process,
under the influence the pair interaction (1), occurs, on average,
through the same stages encountered in nature, or even in a more
sophisticated atomic MD simulation with {\em ab initio} force fields.
Certainly, there are situations where the influence of the
native-state topology on the folding process may be overridden by
strong chemical propensities to form definite pairs of amino acids
(such as disulfide bridges).  In addition, given the explicit bias
towards the native state, one should not expect that it would be
possible to observe intermediate states with low concentration of
native contacts.  Aside from similar circumstances, it is appropriate
to ask whether one can reproduce the key steps of the folding process
by exploiting only the structural information of the native state.
The basis and justification for the present study is the growing
evidence that the above question has a positive answer.  In fact,
starting from the work of ref.~\cite{Micheletti99b} and later of
refs.~\cite{Finkel,eaton,baker,dobson,serrano,Clem}, it has become
clear that the characterization of the transition states can be
confidently done within a Go-model scheme for a variety of proteins.

\begin{figure}
\begin{center}
\psfig{figure=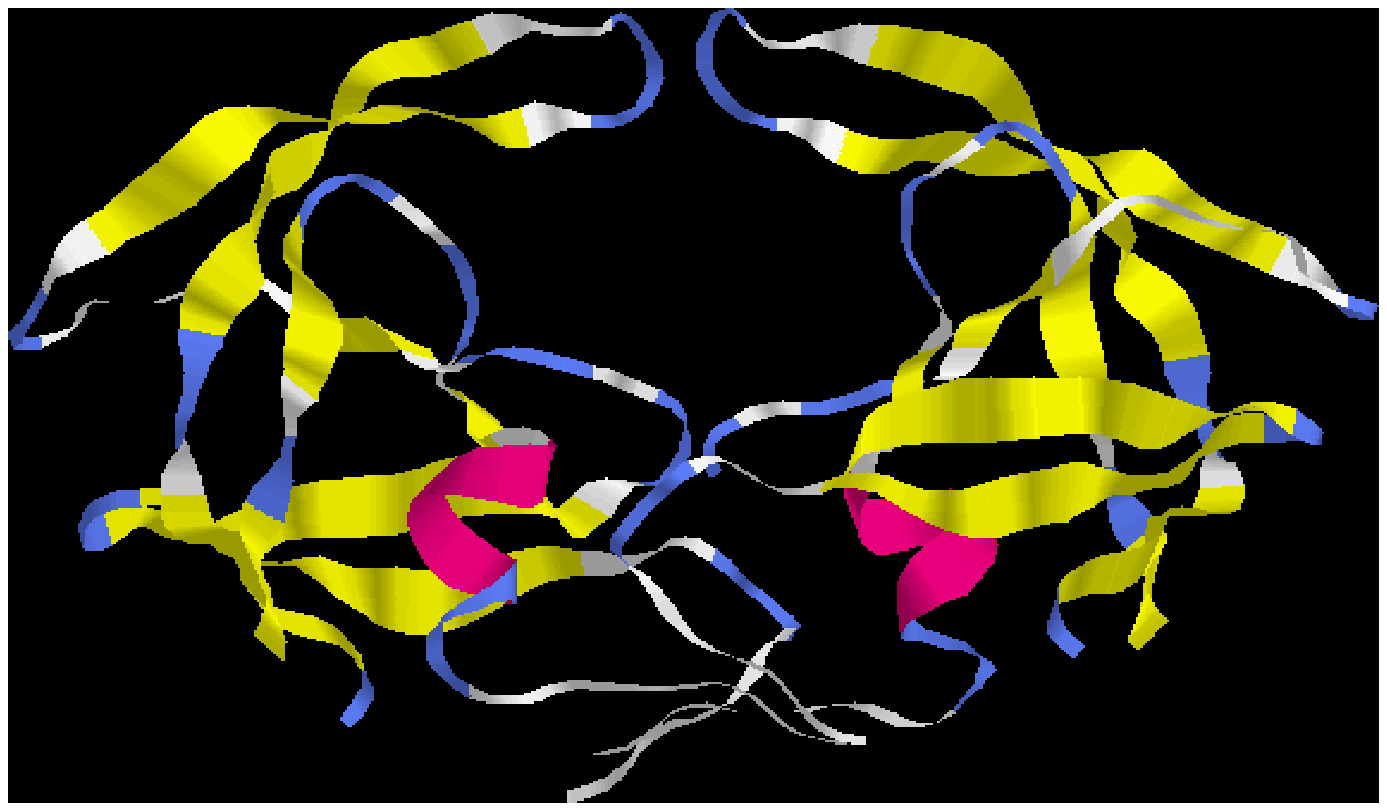,width=\textwidth}
\end{center}
\caption{Structure of HIV-1 PR \protect\cite{condra}.}
\label{fig:dimero}
\end{figure}

In the present study, the starting structural model (target) is the
free enzyme \cite{condra} which is a homodimer with each subunit 
composed by 99 residues, see Fig~\ref{fig:dimero}. 
Following~\cite{CCM}, the crystallographic C2 symmetry was enforced during 
MD simulations to reduce the computational
effort.  The progress towards the fully folded (native) state was
estimated in terms of the fraction of native contacts formed at any
given time in the partially folded structure, $\Gamma$
\cite{reaction}.  
This quantity, also termed overlap, is defined as 
\begin{equation}
Q = \frac{\sum_{i,j} \varepsilon^N_{ij} \cdot \varepsilon^\Gamma_{ij}}
{\sum_{i,j} \varepsilon^N_{ij}}\;, 
\label{eq:ovlp}
\end{equation}

where $\varepsilon^\Gamma$ is the contact matrix of $\Gamma$.

Constant temperature MD simulations were carried out for several
decreasing temperatures from the unfolded to folded state of the
protein.  The equations of motion for the $C_\alpha$ atoms were
integrated by a velocity-Verlet algorithm with time step $\Delta t
=0.01$ combined with the standard Gaussian isokinetic scheme
\cite{Therm}.  We performed unfolding simulations within the same
framework by starting from the native structure and taking it through
a sequence of increasing temperatures (heat denaturation).  The
temperature was measured in reduced units V$_0$/k$_B$ (being k$_B$ the
Boltzmann constant, and V$_0$ the energy of the native contacts in
Eq.~(\ref{eq:inter})).  At each temperature, we let the system
equilibrate from the last structure reached at the previous run at a
nearby temperature.  Each equilibration involved $5 \cdot 10^5$ MD
steps, a time much longer than the largest correlation times observed
for the system.  After equilibration, we sampled $4000$ structures
again at time intervals twice the estimated correlation time.  At each
temperature, we collected the energy histogram of such uncorrelated
structures. Using the multiple histogram techniques \cite{ferren},
the energy measurements for all temperatures have been reweighted to
provide optimal estimates of thermodynamic quantities such as the
average energy and the specific heat (by differentiation of the
former) for a continuous range of temperatures.  The statistical
significance of the data collected in our runs was checked by
verifying that the reweighted thermodynamic quantities did not change
by more than a few percent upon addition of energy histograms obtained
from folding/unfolding simulations with different temperature
schedules or initial conditions.
 
Within the approximation where the reaction coordinate is the internal
energy, from the high temperature side, the slowest dynamics occurs at
temperatures near the specific heat peak, with a relaxation time at
least of order $D \cdot TC_{V}$, where $D$ is a suitable coefficient
dimension of seconds/Joules. Thus the contacts contributing more to
the specific heat peak are identified as the key ones belonging to the
folding bottleneck and sites sharing them as the most probable to be
sensitive to mutations.  Furthermore, by following several individual
folding trajectories (by suddenly quenching unfolded conformations
below the folding temperature, $T_{fold}$) we ascertained that all
such dynamical pathways encountered the kinetic bottlenecks described
in the next section.

\begin{figure}
\begin{center}
\psfig{figure=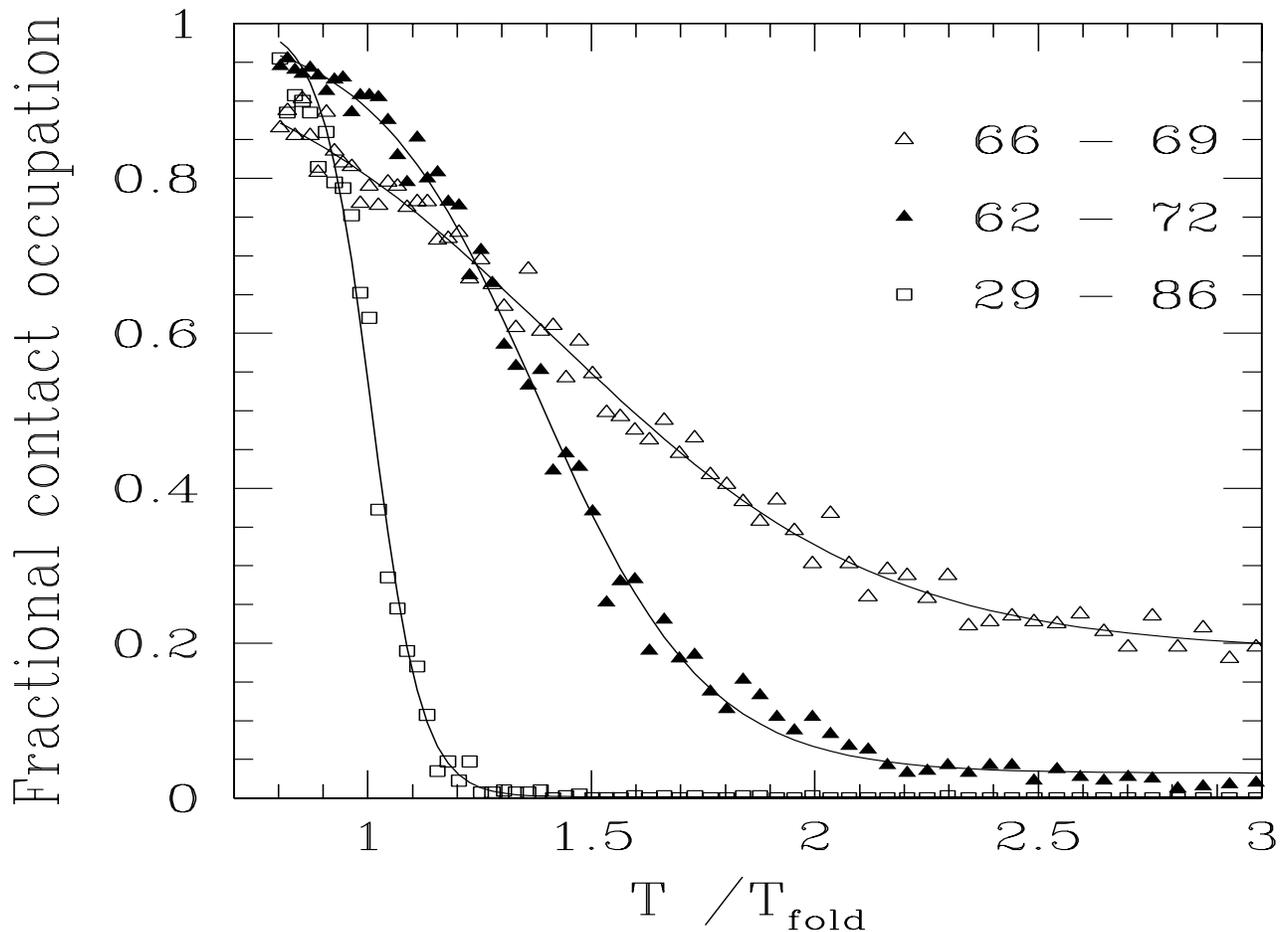,width=\textwidth,height=13.0cm}
\end{center}
\caption{Fractional occupation for three different native contacts.
In all cases the fractional occupation approaches 1 as $T \to 0$ and
vanishes for very large $T$. However, the rapidity of formation (slope
at the inflexion point) is very different. The contact binding
residues 66 and 69, located at the turn of $\beta$ sheet 1, forms very
gradually. The highest rapidity of formation of contacts in $\beta_1$
is observed for the pair 62, 72 (solid triangles). At the folding
temperature, one of the highest formation rapidities is found in
correspondence of the contact bonding residues 29 and 86 (open
squares). The continuous curve is obtained from a smooth interpolation
of the points.}

\label{fig:fit}
\end{figure}

A reliable and convenient way to identify and characterize the kinetic
bottlenecks is through the location of peaks and shoulders in the
specific heat (which denote the overcoming of free energy barriers).
Moreover, since the specific heat results from the contribution of
each pair of native contacts, it is also useful to monitor the
formation of each native interaction throughout the folding process.
Indeed, the probability of formation of a native contact is a
decreasing function of $T$ and has a sigmoidal shape fitted by
suitably shifted hyperbolic tangent (see Fig.~\ref{fig:fit}) . The
smooth interpolation allows to identify a crossover temperature,
$T_0$, where the slope reaches its maximum, $C_0$ (that is the
inflection point of the curves in Fig.~\ref{fig:fit}). $T_0$ defines a
local ``transition'' temperature at which each contact is locked,
whereas $C_0$, which represents the ``rapidity'' of its formation, can
also be regarded as a measure of the local contribution to the
specific heat.

\section{Results}

\begin{figure}
\begin{center}
\psfig{figure=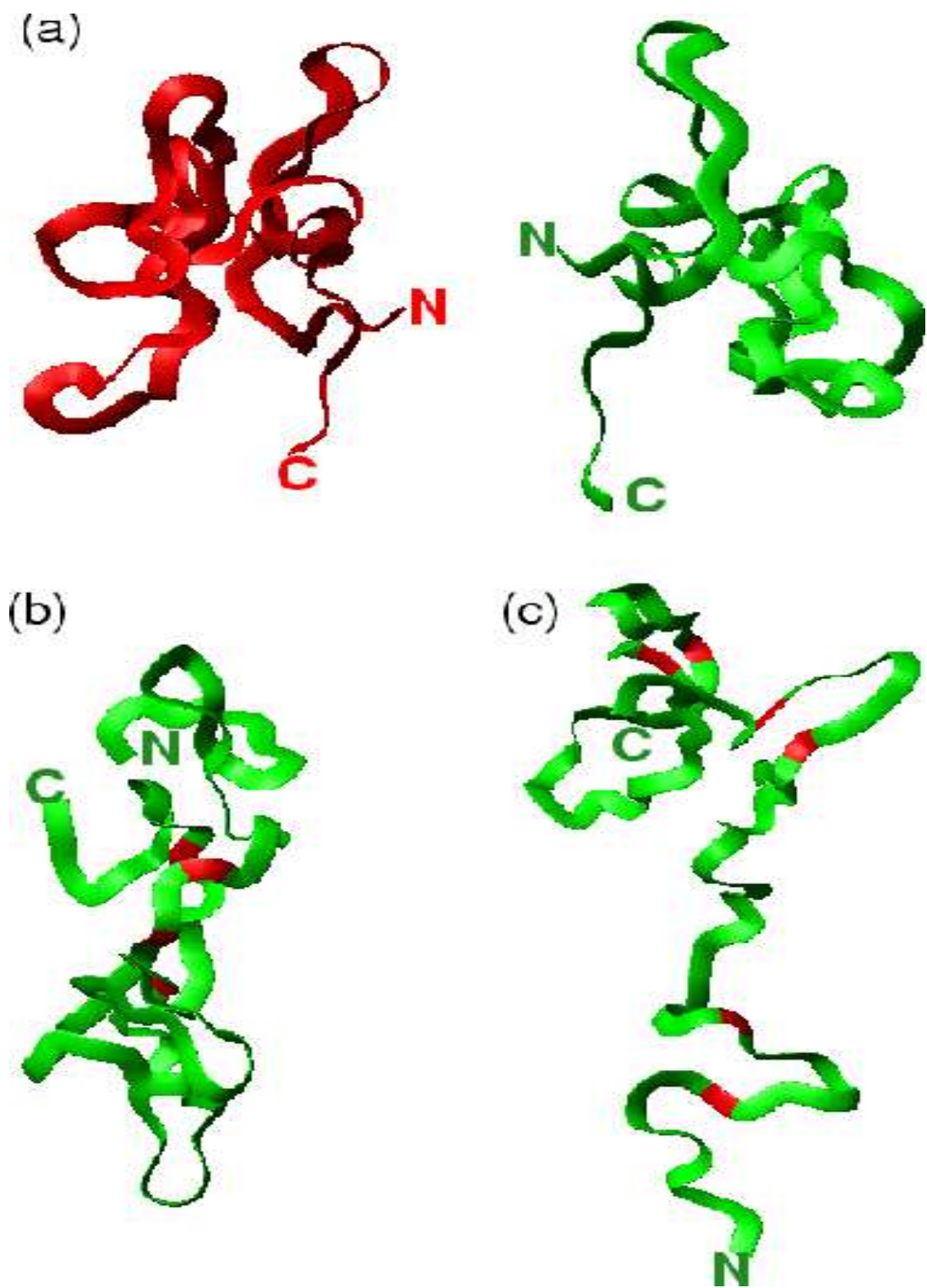,width=0.8\textwidth,height=20.0cm}
\end{center}
\caption{Typical dimer conformations near the dissociation temperature
$T_{diss}$ (a). Typical monomer structures at the folding transition,
$T_{fold}$ (b) (key residues 29, 32, 76, 86 are highlighted in red)
and at $1.3 T_{fold}$ (c) (sites 11, 21, 46, 55, 61, 74 responsible
for the initiation of the beta sheets are shown in red.)}
\label{fig:diss}
\end{figure}

\begin{figure}
\begin{center}
\psfig{figure=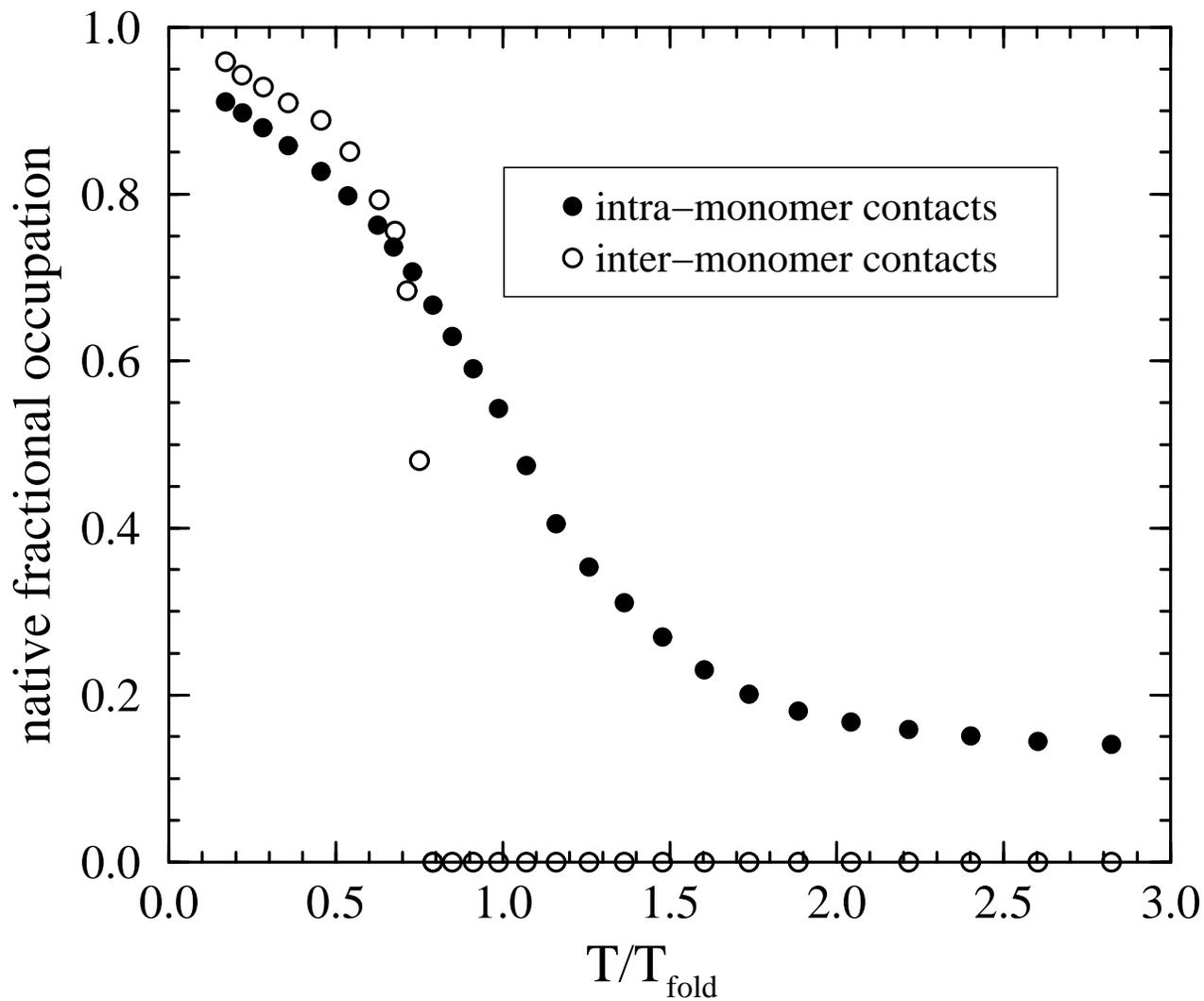,width=0.8\textwidth,angle=270}
\end{center}
\caption{Behaviour, as a function of temperature, of the average
fractional occupation of native contacts within each HIV-1 PR monomer
(solid circles) and at the interface between the two monomeric units
(open circles). The dissociation of the two subunits is clearly seen
to occur at $T/T_{fold}=0.6$.}
\label{fig:ovlp}
\end{figure}

\begin{figure}
\begin{center}
\psfig{figure=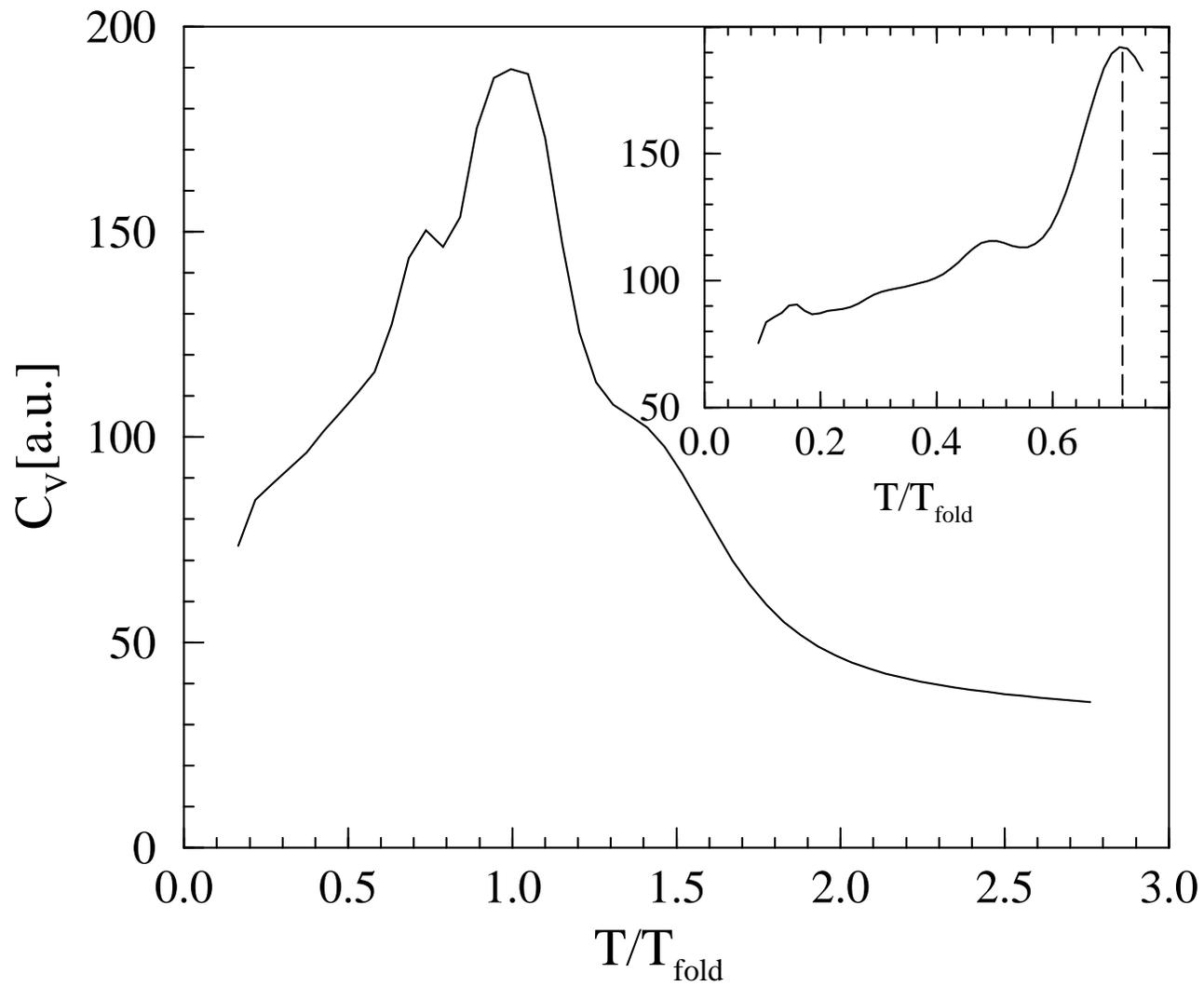,width=0.8\textwidth,angle=270}
\end{center}
\caption{Specific heat of the HIV-PR single monomer.  The dimer
specific heat at temperatures below the $T_{diss}$ is shown in the inset.}
\label{fig:specheat}
\end{figure}

At very low temperatures, the observed structures have nearly 100 \%
native-state similarity, measured as the fraction $Q$ of established
native contacts (Eq.~\ref{eq:ovlp}).  Further increase in temperature
causes structural rearrangements into a configuration that cannot be
assembled into a dimer anymore (Fig.~\ref{fig:diss}): the number of
subunit-subunit contacts vanished and the two subunits behaved
independently.  The dissociation mechanism is well described by
Fig.~\ref{fig:ovlp} where we report, for several temperatures, the
fractional occupation of native contacts for the individual subunits
and at the monomer-monomer interface. The dissociation is also
signalled by an abrupt increase of the specific heat of the dimer, see
inset of Fig.\ref{fig:specheat}, (this defines the the dissociation
temperature $T_{diss}$).  A typical structure at this temperature is
shown in Fig. \ref{fig:diss}a. At even higher temperatures ($T = 1.4
T_{diss}$), a large increase of specific heat is observed, indicating
the presence of a strong transition of the single subunits
\cite{scheraga}, see Fig.~\ref{fig:specheat}. This temperature is
identified with the folding temperature, $T_{fold}$. Consistently
with other studies on different proteins, the native overlap at
$T_{fold}$ was about 50 $\%$. A typical structure at this temperature
is shown in Fig.  \ref{fig:diss}b.
 
A further set of bottlenecks is encountered at $T \approx 1.4
T_{fold}$, where the formation of the three $\beta$-strands of HIV-1
PR is involved. Upon increasing the temperature one encounters
$\beta$-sheet $\beta_2$, then $\beta_1$, then $\beta_3$.  It is found
that the kinetic bottleneck for the a general $\beta$-sheet formation
is not the establishment of the contact(s) closest to the $\beta$ turn
(that involves amino acids near in sequence) but it is located further
away.  A quantitative analysis of the amino acids most involved in the
folding bottleneck is again obtained by monitoring during the
folding/unfolding process each pair of amino acids which are in
contact in the native state. Examples of the probability with which
individual contacts are formed is shown in Fig.~\ref{fig:fit}.

\begin{figure}
\begin{center}
\psfig{figure=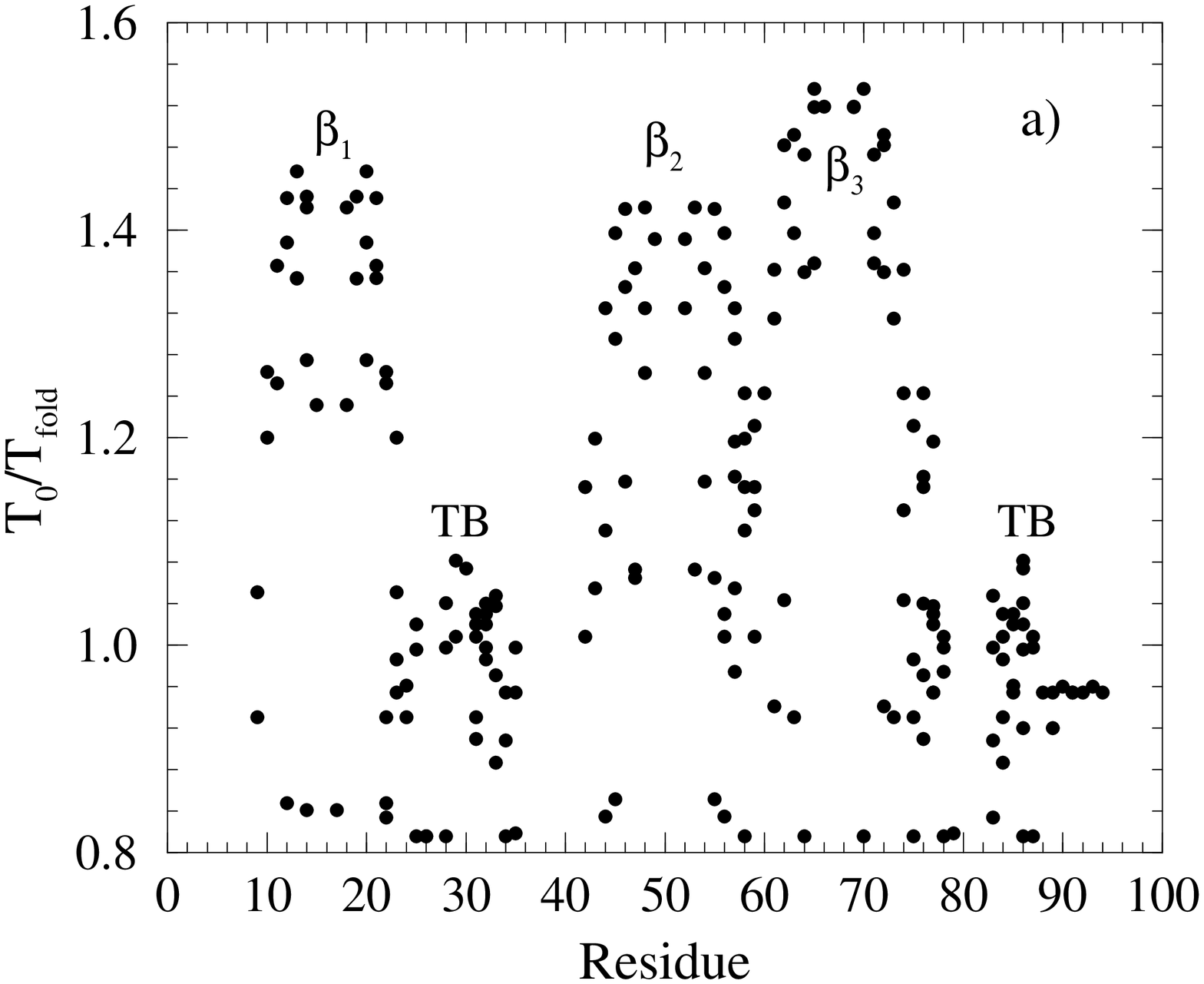,width=2.8in,height=3.5in,angle=270} 
\psfig{figure=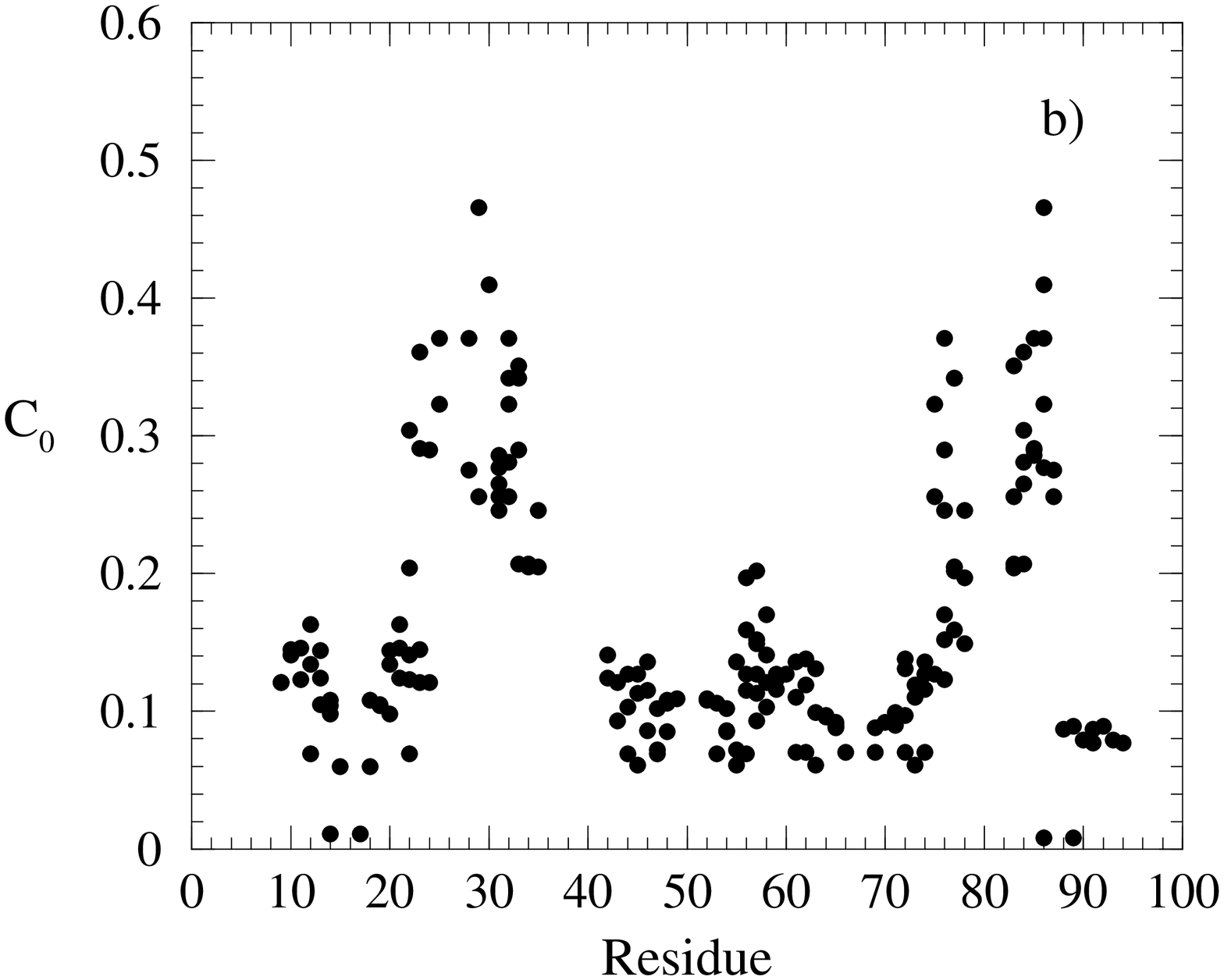,width=2.8in,height=3.5in,angle=270}
\end{center}
\caption{(a) Characteristic temperatures $T_0$ (a) and maximum
``rapidity'' $C_0$ associated to each native contact versus the amino
acid position sharing the contact. (b) Distribution of the values
of the maximum rapidity $C_0$ of contacts involving each residue.}
\label{fig:flessi}
\end{figure}

At each temperature where the dynamical evolution of the HIV-1 PR is
followed, the formation probability of each native contact (fractional
occupation) is calculated.  Such quantities are $1$ at very low
temperatures (all native contacts always present) and decrease to zero
at temperatures larger than the folding temperature. It may be
anticipated that the rate of decrease as a function of temperature
will not be the same for all contacting pairs. In particular, trivial
local contacts between residues with a small sequence separation will
have a large probability of being formed even at high temperatures.
Our interest focuses on those contacts which show a dramatic increase
of the fractional occupation near the folding transition. Those will
be the key contacts responsible for the appearance of the specific
heat peak. Examples of the fractional occupation for three native
contacts is shown in Fig.~3. Given the monotonic behaviour of the
fractional occupation, one may synthetically characterize the
formation of each contacting pair by the temperature at which the
point of inflection of the curve is seen and also by the slope at that
very same point. Both data can be conveniently summarised in two
scatter plots where the slope, $C_0$, and the temperature of
formation, $T_0$, are reported for each residue taking part in native
contacts. Such graphs are reported in the scatter plot of
Fig. \ref{fig:flessi}.  Notice that, for each site, there are as many
entries as the number of contacts involving it (a number that
typically differs from site to site).  Figure \ref{fig:flessi} clearly
shows that there are clusters of contacts that are turned on at
similar temperatures.

\begin{figure}
\begin{center}
\psfig{figure=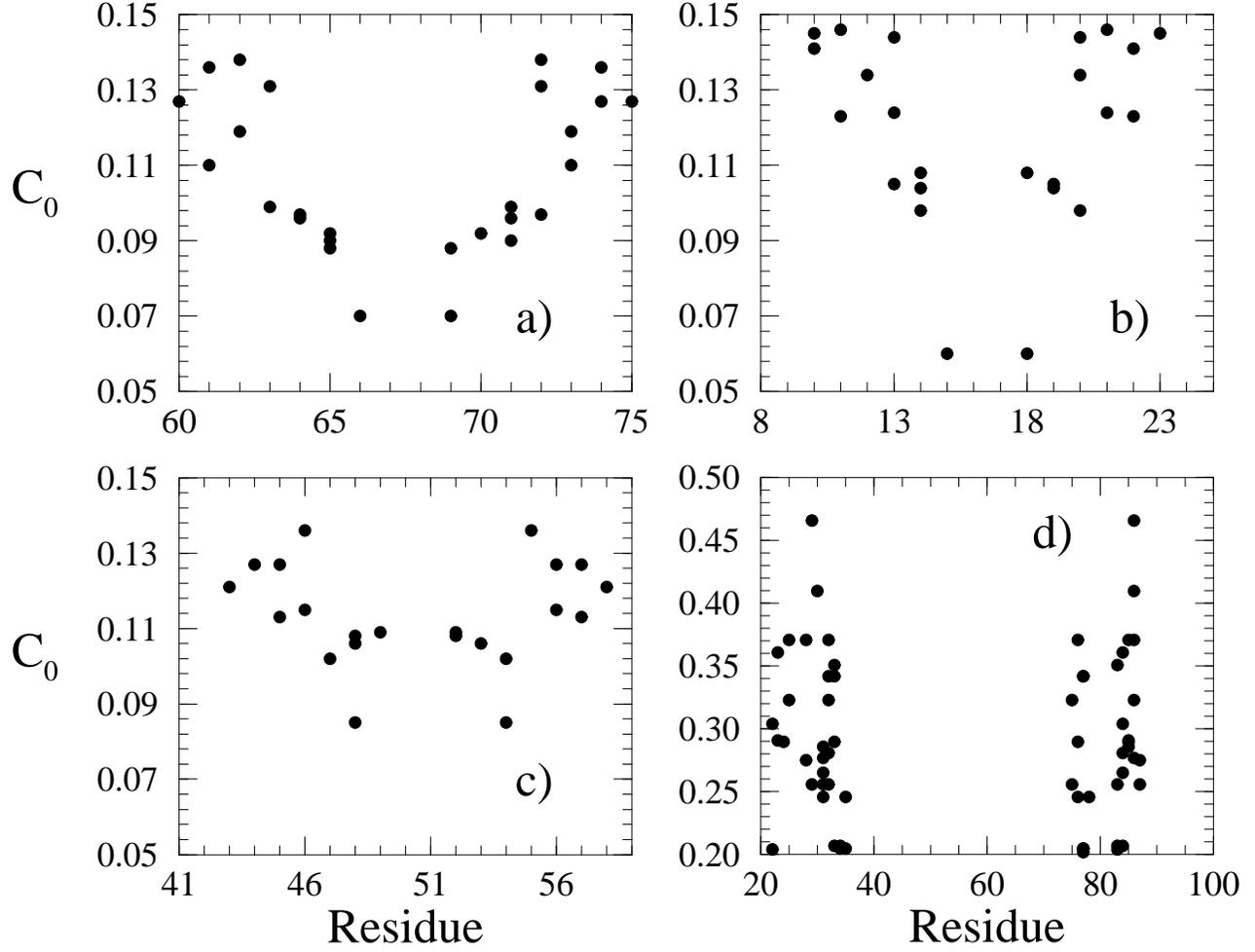,width=0.75\textwidth,angle=270}
\end{center}
\caption{Formation ``rapidity'' $C_0$, for contacts in the four
subregions of Fig.~\ref{fig:flessi}a. In order of decreasing formation
temperature data are shown for a) $\beta_3$, b) $\beta_1$, c) $\beta_2$
and d) TB. Notice that the scale of $C_0$ depends on temperatures. The
highest $C_0$'s for each of the regions increase as the temperature
$T_0$ decreases.}
\label{fig:pendenze}
\end{figure}

The bottlenecks for the folding process are identified by isolating
the contacts having both a formation temperature, $T_0$ matching the
location of the peaks and shoulders of the specific heat, and a high
rapidity of formation, $C_0$.  Figure \ref{fig:flessi}a reveals the
presence of four distinct clusters of contacts.  The first three,
labelled $\beta_1$, $\beta_2$, $\beta_3$, are associated with the
formation of the three antiparallel $beta$-sheets in HIV-1 PR.  Their
temperature of formation is about $1.4 \cdot T_{fold}$, and
corresponds to the shoulder visible in the larger plot of the specific
heat of Fig.~\ref{fig:specheat}.  The sites sharing the most important
contacts involved in such three bottlenecks are listed in
Table~\ref{tab:tab} and highlighted in Fig. \ref{fig:diss}c, where a typical
structure at $T = 1.4 \cdot T_{fold}$ is shown.  It is interesting to
see that $T_0$ is maximum for sites close to the $\beta$-turn, in
accord with the intuitive expectation that the $\beta$ formation is
initiated at the turn.  In Fig. \ref{fig:pendenze} a,b and c, we have 
reported the
values of $C_0$ only for the pairs of contacting sites in the $\beta$
sheet.  It is seen that the sites closest to the turn have a small
formation rapidity. This can be understood since, being very close
along the sequence, they can be easily formed/broken. The highest
rapidity, $C_0$, i.e. the highest difficulty of formation, is
encountered typically 3-4 sites away from the turn. The corresponding
contacts are then identified as the bottleneck for this particular
folding stage. For the $\beta$-sheets, the bottlenecks involve amino
acids that are typically 3-4 residues away from the turns.

Going back to Fig.~5, one observes that there is a fourth group of
contacts around residues 30 and 86, labelled TB after ``transition
bottleneck'', that are formed cooperatively at the folding
transition. The sites involved in the TB contacts are listed in 
Table~\ref{tab:tab}.    
Among those contacts we have recorded the largest values
of $C_0$, as shown more clearly in Fig. 6d. Again, we considered the
sites with the highest values of $C_0$ as responsible for the main
bottleneck of the folding process.  The highest ``rapidity'' is
measured in correspondence of contacts 29-86 and 30-84 (see also
Fig.~\ref{fig:pendenze}) which are, consequently, identified as the
most crucial for the folding/unfolding process. 

\section{Discussion}

The sites involved in the main folding bottleneck (TB) are located at
the active site of HIV-1 PR, which is targeted by anti AIDS drugs
\cite{condra1}. Hence, within the limitations of our simplified
approach, we predict that changes in the detailed chemistry at the
active site affect also a key step of the folding process. To
counteract the drug action, the virus has to perform some very
delicate mutations in correspondence of the key sites; within a random
mutation scheme this requires many trials (occurring over several
months). The time required for the biosynthesis of a mutant with
native-like activity is even longer if the drug attack correlates with
several bottlenecks simultaneously.

This is certainly the case for several anti-AIDS drugs. Indeed Table
~\ref{tab:tab} summarises the mutations emerged for the FDA approved
drugs \cite{BIOCH88}. 
Remarkably, among the first 23 most crucial sites   
predicted by our method and listed in Table~\ref{tab:tab}), there are 7 sites 
in common with the 16 distinct mutating sites of Table \ref{tab:tab}. The
probability that two sets of 16 and 23 sites randomly taken from a
total population of 99 (the length of the HIV-1 PR monomer) share at
least 7 sites is only 3 \%. Also note that, all the mutation sites of
Table \ref{tab:tab} except 82,35,36 and 90 fall within a mismatch of
at most one position from the sites of Table \ref{tab:tab2}. These
results highlight the highly statistical correlation between our
prediction and the evidence accumulated from clinical trials.

All mutations causing resistance involve crucial residues for the main
folding bottleneck, (particularly residue 84) in combination with key
sites for one or more of the $\beta$ sheets.  Mutation in this sites
are expected to modify the energetics and structure of partially
folded states. In contrast the folded state appears to be weakly
affected by specific mutations, such as M46I, L63P, V82T, I84V, which
lead to a $C_{\alpha}$ RMS distance of 0.5 \AA from the wild-type
\cite{Chen,Nair}.  In the light of these results, it is possible to
interpret the experimental evidence for the existence of correlations
between mutations at residue 82 and residues 10, 54, 71 as correlations
between the main kinetic barrier, TB and the others $\beta_1$,
$\beta_2$, $\beta_3$. The large separation of these associated sites,
both along the sequence and in space, suggests that their correlations
arise by virtue of the folding process itself. This kinetic effects is
particularly clear in one of these cases, namely the co-mutation of
sites in the TB and at residue 10, which occurs under IND therapy. The
mediator of the correlation is residue 23 which takes part to two
bottlenecks: TB and $\beta_1$ through direct contact with residues 84
and 10, respectively. Co-varying mutations between the two sites are
observed because changes in TB will affect the environment of the other
key site 10, which has to mutate accordingly.

\section{Conclusions}

The strategy presented here allows both to identify the bottleneck of
the folding process and to explain their highly significant match with
known mutating residues. This approach should be readily 
applicable to identify the kinetic bottlenecks of other viral enzymes 
of pharmaceutical interest, thus aiding the development of novel inhibitor
targetting the kinetic bottlenecks. This is expected to enhance
dramatically the difficulty for the virus to express mutated proteins
which still fold efficiently into the same native state with unaltered
functionality.
          
{\bf Acknowledgements} This work was supported by INFM and MURST.


\newpage

\newpage

\newpage
\vspace{1.cm}
\newpage
\vspace{1.cm}
\newpage
\newpage

\newpage

\begin{table}
\begin{center}
\begin{tabular}{|l|l|}
Bottleneck & Key sites \\ \hline \hline
TB & 22, 29, 32, 76, 84, 86 \\
$\beta_1$ & 10,11,13,20,21,23 \\
$\beta_2$ & 44,45,46,55,56,57 \\
$\beta_3$ & 61,62,63,72,74 \\
\end{tabular}
\end{center}
\caption{Key sites for the four  bottlenecks. For each bottleneck, only the
sites in the top three pairs of contacts have been reported.}
\label{tab:tab}
\end{table}
\begin{table}
\begin{center}
\begin{tabular}{|l|l|l|} 
Name & Point Mutations & Bottlenecks \\ \hline \hline 
RTN \protect\cite{Molla,Marko} & {\bf 20}, { 33}, 35, 36, {\bf
46}, 54, {\bf 63}, 71, 82, {\bf 84}, 90 & TB, $\beta_1$, $\beta_2$, $\beta_3$ \\
NLF \protect\cite{patick} & 30, {\bf 46}, {\bf 63}, 71, { 77}, {\bf 84}, 
& TB, $\beta_2$, $\beta_3$ \\
IND \protect\cite{condra,tisdale} & {\bf 10}, {\bf 32}, {\bf 46}, {\bf 63},71, 82,
{\bf 84} & TB, $\beta_1$, $\beta_2$, $\beta_3$ \\
SQV \protect\cite{condra,tisdale,jacob} & {\bf 10}, {\bf 46}, 48, {\bf 63}, 
71, 82, {\bf 84}, 90 & TB, $\beta_1$, $\beta_2$, $\beta_3$ \\
APR \protect\cite{apr} & {\bf 46}, {\bf 63}, 82, {\bf 84} & TB, $\beta_2$, $\beta_3$ \\
\end{tabular}
\end{center}
\caption{Mutations in the protease associated with FDA-approved drug
resistance \protect\cite{BIOCH88}. Sites highlighted in boldface are those involved in the
folding bottlenecks as predicted in our approach. $\beta_i$ refers to
the bottleneck associated to the formation of the $i$-th
$\beta$-sheet, whereas TB refers to the bottleneck occurring at
folding transition temperature $T_{fold}$}
\label{tab:tab2}
\end{table}

\end{document}